\documentclass[preprint,12pt,rmp]{revtex4}
\usepackage{epsfig}
\usepackage{graphicx}
\usepackage{dcolumn}
\usepackage{bm}
\usepackage{ulem}
\usepackage{amsmath}

\begin{document}


\title{Asymmetry in the presence of migration stabilizes multistrain disease outbreaks}

\author{Simone Bianco}
\author{Leah B. Shaw}
\affiliation{Department of Applied Science, The College of William and Mary,
  PO Box 8795, Williamsburg, VA, 23187-8795}

\date{\today}
\begin{abstract}
  We study the effect of migration between coupled populations, or patches, on the stability
  properties of multistrain disease dynamics. The epidemic model used in
  this work displays a Hopf bifurcation to oscillations in a single well
  mixed population. It
  is shown numerically that migration between two non-identical patches stabilizes
  the endemic steady state, delaying the onset of large amplitude outbreaks
  and reducing the total number of infections.
  This result is motivated by analyzing generic Hopf bifurcations with
  different frequencies and with diffusive coupling between them.
  Stabilization of the steady state is again seen, indicating that our
  observation in the full multistrain model is based on qualitative
  characteristics of the dynamics rather than on details of the disease
  model.
\end{abstract}
\maketitle

\section{Introduction}
In this work we study the stability of
a multistrain disease model in two coupled populations. Multistrain diseases are diseases
with multiple coexisting strains, such as
influenza \citep{AndreasenLL97}, HIV~\citep{Hu1996}, and dengue \citep{FergusonDA99}.
We consider two kinds of strain
interactions: cross immunity and antibody-dependent enhancement. When the
disease infects an individual, his or her immune system creates serotype-specific
antibodies, which will protect the individual against that
serotype. However, antibodies may also give some cross-protection to the
other serotypes~\citep{Hal2007}. This reduced susceptibility to the other serotypes may be temporary. When the temporary cross immunity wanes,
heterologous secondary infections are possible. Low level antibodies developed
from primary infections may form complexes with the virus so that more cells are
affected, and viral load is increased \citep{Vaughn00}. This effect is called
antibody-dependent enhancement (ADE),
and it has been observed in vitro in diseases such as ebola~\citep{TakEtAl03}
and dengue~\citep{HalORo77}. Throughout this work we make the hypothesis that ADE increases the
infectiousness of secondary infective cases due to the higher viral
load~\citep{LeahPNAS,LeahPRE}. Alternative views of ADE as an increase
in mortality associated
with secondary infectives can be considered~\citep{KawSasBoo03}.
We focus in this paper on the multistrain disease dengue, which is
believed to exhibit both temporary cross immunity and
antibody-dependent enhancement \citep{Hal2007}. Dengue is a
subtropical
mosquito-borne disease that exhibits up to four serotypes.  It is
widespread in tropical regions of southeast Asia, Africa, and the
Americas, infecting an estimated 50 to 100
million people every year~\citep{B:who1}. Primary infections may be asymptomatic,
while secondary infections are more severe, with about $5\%$ of secondary
infections leading to dengue hemorrhagic fever (DHF) or dengue shock syndrome
(DSS), the potentially fatal forms of the disease \citep{B:who1}.
  An effective vaccine against dengue is very difficult to achieve.
Because
  of ADE, infection with an unvaccinated strain following a single strain vaccination may lead to the more severe symptoms
 associated with secondary
infections~\citep{HalDee02}. 
 Therefore, an effective vaccine must
protect against all four serotypes simultaneously.
 A recent
theoretical study~\citep{LoraIra08} has shown that eradication of
dengue using only single-strain vaccines is unlikely. Because a
tetravalent vaccine is not immediately forthcoming, deepening our
understanding of the dynamics of dengue in more realistic models is of
great importance.

The dynamics of dengue in a single, well mixed population has been studied
in several recent publications, including but not limited to \citet{FerAndGup99,LeahPRE, LeahPNAS, WeaRoh06, BiaShaSch09}.
ADE and cross immunity have been
shown to play a fundamental role in the spreading of dengue, causing
instability, desynchronization of serotypes, and chaotic
outbreaks~\citep{LeahPRE,LeahPNAS,BiaShaSch09}.  However, real
populations may be spatially heterogeneous.  To gain further insight
into the wide spectrum of possible dengue dynamics, we relax the
assumption of a well mixed population.    The division of a population
into spatially distinct patches simulates the potentially heterogeneous
environment in which
the disease spreads. Spatial heterogeneity has been
invoked in the past to account for the persistent character of some infectious
diseases~\citep{LloMay96, HagFer04} and to explain in-phase and
out-of-phase dynamics of diseases~\citep{HeSto03}. A multipatch model with age structure has
been used to explain bi- and tri-annual oscillations in the spread of
measles~\citep{BolGren95}.

Including human mobility patterns has a significant influence on the spreading of infectious
disease. We will assume here that the coupling between patches is via migration, movement of
individuals from one patch into another (as in \citet{LieSch04,RuaWanLev06,
SatDie95,GraEllGla03}). An alternative coupling strategy is mass
action coupling, in which susceptibles in one patch are assumed to
interact with infectives in another patch, introducing nonlinear
coupling terms \citep{BolGren95, LloMay96, RohEtAl99}.
Stochastic coupling has also been modeled (see \citet{KeeRoh02} and references therein, and~\citet{ColVes08}).

The purpose of this work is to analyze the stability of a model for
multistrain diseases with interacting strains, using dengue as an example, in a system of two coupled
patches.  Since chaotic outbreaks are likely to produce a
  higher number of infected individuals, understanding the stability properties
  may play an important role in public health. The case of non-identical parameters in the two patches will
be of particular interest, as this serves as a model for spatial
heterogeneity.  The paper is divided as follows. In
Section~\ref{sec:model} we introduce the epidemic model.  Section
\ref{sec:1patch} summarizes the bifurcation structure for a single
patch.  In Section~\ref{sec:2patch} we present numerical results for
bifurcations in two coupled patches.  Section~\ref{sec:analytical}
motivates these results via analysis of a simple, lower dimensional
model,
and Section~\ref{sec:conclusions} concludes.

\section{Model}
\label{sec:model}
We use a compartmental model for multistrain disease spread with cross
immunity and antibody-dependent enhancement, which was previously studied in a
single well-mixed population by \citet{BiaShaSch09}.  In this model, individuals
may develop a primary infection with any of the serotypes.  Immediately after
recovering, the individual experiences a period of temporary partial cross
immunity to all other serotypes.  When the cross immunity wears off,
immunity to the primary infecting strain is retained, but the individual may
develop a secondary infection with a different serotype.  Infectiousness of
the secondary infectives is increased due to antibody-dependent enhancement.
After the secondary infection, complete immunity to all serotypes is assumed.
A flow diagram for the single patch model with two serotypes is shown in Figure
\ref{fig:flowdiagram} for simplicity, but we present results here for all four
serotypes.  We extend the model to two spatially distinct patches,
which are coupled by linear migration terms. For two patches (indexed by $q$)
and $n$ strains ($n$ arbitrary), the model is as follows:
  \small
  \begin{eqnarray}
    \frac{ds_q}{dt} & = & \mu_q - \beta_q s_q \sum_{i = 1}^{n} \left( x_{q,i} + \phi \sum_{j
      \neq i} x_{q,ji} \right) - \mu_{q} s_q  \nonumber \\
      && - \nu s_q+\nu s_{q'} \label{odefirst}
  \end{eqnarray}
  \begin{eqnarray}
    \frac{dx_{q,i}}{dt} & = & \beta_q s_q \left( x_{q,i} + \phi \sum_{j \neq i} x_{q,ji}
    \right) - \sigma x_{q,i} - \mu_{q} x_{q,i} \nonumber \\
    && -\nu x_{q,i}+\nu x_{q',i}
  \end{eqnarray}
  \begin{eqnarray}
    \frac{dc_{q,i}}{dt} & = &  \sigma x_{q,i} - \beta_q (1-\epsilon) c_{q,i}  \sum_{j
      \neq i}
    \left( x_{q,j} + \phi \sum_{k \neq j} x_{q,kj} \right) \nonumber \\
          && - \theta c_{q,i} - \mu_q c_{q,i}  -\nu c_{q,i}+\nu c_{q',i}
  \end{eqnarray}
  \begin{eqnarray}
    \frac{dr_{q,i}}{dt} & = &  \theta c_{q,i} - \beta_q r_{q,i} \sum_{j \neq i}
    \left( x_{q,j} + \phi \sum_{k \neq j} x_{q,kj} \right) - \mu_{q} r_{q,i} \nonumber \\
    && -\nu r_{q,i}+\nu r_{q',i}
  \end{eqnarray}
  \begin{eqnarray}
    \frac{dx_{q,ij}}{dt} & = &  \beta_q r_{q,i} \left(
    x_{q,j} + \phi \sum_{k\neq j} x_{q,kj} \right) \nonumber \\
    &&+ \beta_q (1-\epsilon) c_{q,i} \left( x_{q,j} + \phi \sum_{k\neq j} x_{q,kj}
    \right)
     -\sigma x_{q,ij}  \nonumber \\
     && - \mu_{q} x_{q,ij} -\nu x_{q,ij}+\nu x_{q',ij} \label{odelast}
  \end{eqnarray}
\normalsize
where the variables are $s_q$, the fraction of susceptibles in patch
$q$; $x_{q,i}$, the fraction of primary infectives with strain $i$ in
patch $q$; $c_{q,i}$, the fraction of individuals in patch $q$ with
partial cross immunity to strain $i$; $r_{q,i}$, the fraction of
individuals in patch $q$ that are recovered from a primary infection
with strain $i$ and no longer have cross immunity to the other
strains; and $x_{q,ij}$, the fraction of individuals in patch $q$ that
were previously infected with strain $i$ and currently infected with
strain $j$.   The parameters are the number of strains $n$, the
contact rate in patch $q$ $\beta_q$,
the recovery rate $\sigma$, the ADE factor $\phi$, the strength of cross
immunity $\epsilon$, the rate $\theta$ for cross immunity to wear off, the
birth and mortality rate in patch $q$ $\mu_q$, and the migration rate
between patches $\nu$. A list of parameters appears in
Table~\ref{tab1}.

For simplicity, the birth and death rates in a patch are set equal to
each other so that the total population of each patch is constant. The
model of Eqs.~\ref{odefirst}-\ref{odelast} allows for one reinfection.
The parameter $\epsilon$ determines how susceptible the cross immune compartments
$c_i$ are to a secondary infection, where $\epsilon=0$ means
no cross immunity (the infection rate is identical for compartments
$c_i$ and $r_i$) and $\epsilon=1$ confers complete cross
immunity (cross immunes are immune to a secondary infection for an average time $\theta^{-1}$). We
allow $\epsilon$ to take any value between $0$ and $1$. The ADE factor
$\phi$ is the enhancement in the infectiousness of secondary infectives. $\phi =
1$ means that secondary infectives are as infectious as primary infectives,
while $\phi = 2$ means that secondary infectives are twice as infectious as
primary, and so forth. In contrast to \citet{AguSto07}, we will not consider
values of $\phi$ smaller than $1$.   The migration rate from patch $q$ to $q'$, and from
patch $q'$ to $q$,  is $\nu$.  All individuals migrate with equal
probability, independent of their infection status.  The migration rate is
assumed to be slow compared to the infection spread.  For convenience, we put
it on the same order as the birth/death rate.  We assume that the social
parameters, which are the contact rate $\beta_q$ and birth/death rate $\mu_q$,
may vary between patches, while the epidemic parameters are the same in all
regions. Because the social parameters depend on human factors (and in the
case of the contact rate also include mosquito levels, which are weather-dependent),
these parameters are the most likely to be different in adjacent
regions.  We use parameter values compatible with dengue fever, which
are summarized in  Table~\ref{tab1}.  Our contact rate $\beta$
  corresponds to a reproductive rate of infection $R_0$ of $3.2 -
  4.8$, which is consistent with previous
  estimates~\citep{FergusonDA99, NagKoe08}.


  \setlength{\unitlength}{1cm}
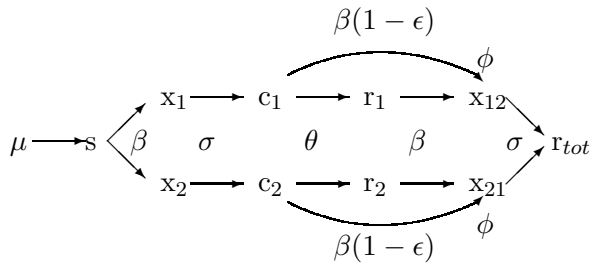
\begin{figure}[!ht]
  \begin{picture}(15,3.)
    \put(1,1){$\mu$}
    \put(1.3,1.12){\vector(1,0){0.7}}

    \put(2,1){s}
    \put(2.3,1.12){\vector(1,1){.5}}
    \put(2.3,1.12){\vector(1,-1){.5}}
    \put(2.6,1){$\beta$}

    \put(3,1.6){x$_1$}
    \put(3.,.45){x$_2$}
    \put(3.4,1.7){\vector(1,0){.7}}
    \put(3.4,.55){\vector(1,0){0.7}}
    \put(3.5,1){$\sigma$}
    \qbezier(4.7,2)(5.9,2.6)(7.2,2)
    \qbezier(4.7,.3)(5.9,-.3)(7.2,.3)
    \put(7.3,1.9){\vector(1,-1){0}}
    \put(7.3,.4){\vector(1,1){0}}

    \put(5.3,2.6){$\beta(1-\epsilon$)}
    \put(5.3,-0.4){$\beta(1-\epsilon$)}

    \put(4.3,1.6){c$_1$}
    \put(4.3,.45){c$_2$}
    \put(4.81,1.7){\vector(1,0){0.7}}
    \put(4.81,.55){\vector(1,0){0.7}}
    \put(4.91,1){$\theta$}

    \put(5.7,1.6){r$_1$}
    \put(5.7,.45){r$_2$}
    \put(6.2,1.7){\vector(1,0){0.7}}
    \put(6.2,.55){\vector(1,0){0.7}}
    \put(6.3,1){$\beta$}

    \put(7.1,1.6){x$_{12}$}
    \put(7.1,.45){x$_{21}$}
    \put(7.6,1.7){\vector(1,-1){0.5}}
    \put(7.6,.55){\vector(1,1){0.5}}
    \put(7.6,1){$\sigma$}
    \put(7.2,2.1){$\phi$}
    \put(7.2,-.1){$\phi$}

    \put(8.2,1){r$_{tot}$}
  \end{picture}
  \vspace{.3cm}
\caption{Flow diagram for single patch model with $2$ serotypes \citep{BiaShaSch09}. Note the reduction of susceptibility to a
  secondary infection through the cross immunity factor ($1 - \epsilon$) and the enhancement of
  secondary infectiousness due to the ADE factor $\phi$. Mortality terms for each
  compartment are not included in the diagram for ease of reading.} \label{fig:flowdiagram}
\end{figure}

\begin{table*}
  \caption{Parameters used in the model}\label{tab1}
  \centering
  \begin{tabular}{lcc}
    Parameter & Value & Reference\\
    \hline\\
    $\mu$, birth and death rate, years$^{-1}$ & $\sim 0.02$ &  Ferguson {\it et al.}(1999)\\
    $\beta$, transmission coefficient, years$^{-1}$ & $\sim 200$ & Ferguson {\it et al.}(1999)\\
    $\sigma$, recovery rate, years$^{-1}$ & $50$ & \citet{RIGAUPEREZ1998} \\
    $\theta$, rate to leave the cross & $2$ &
    Wearing and Rohani (2006)\\
    immune compartment, years$^{-1}$ & & \\
    $\phi$, ADE factor & $\geq 1$ & Schwartz \textit{et al.} (2005) \\
    $\epsilon$, strength of cross immunity & $0-1$ & Bianco \textit{et al.} (2009) \\
    $\nu$, migration rate, years$^{-1}$ & $0-0.05$ & - \\
    $n$, number of strains & 4 & -\\
    \hline
  \end{tabular}
\end{table*}

\section{Single patch bifurcation structure}
\label{sec:1patch}

A similar model  to the one of Eqs.~\ref{odefirst}-\ref{odelast} has recently been used to analyze the dynamics of dengue
fever in a single, well mixed, population~\citep{BiaShaSch09}. The ADE
$\phi$ and cross immunity strength $\epsilon$ were varied as
bifurcation parameters. In the absence
of cross immunity ($\epsilon = 0$), ADE alone generates instability, desynchronization, and ultimately chaotic
outbreaks~\citep{LeahPNAS,LeahPRE}. A Hopf bifurcation is observed for
a critical value of the ADE factor $\phi$, above which oscillatory solutions are obtained. Weak cross immunity
stabilizes the system, while strong cross immunity triggers
instability and chaos even in the absence of ADE
\citep{BiaShaSch09}. In the latter case, destabilization occurs via a
Hopf bifurcation for a critical value of the cross immunity
strength $\epsilon$.  At the bifurcation, three identical complex
pairs of eigenvalues of the Jacobian simultaneously become unstable.
Although \citet{BiaShaSch09} discusses the full two parameter
bifurcation structure (in $\epsilon$ and $\phi$), we will consider the
cross immunity and ADE effects separately in the present work.

\begin{figure}[!ht]
  \begin{center}
    \includegraphics[width=9cm,keepaspectratio]{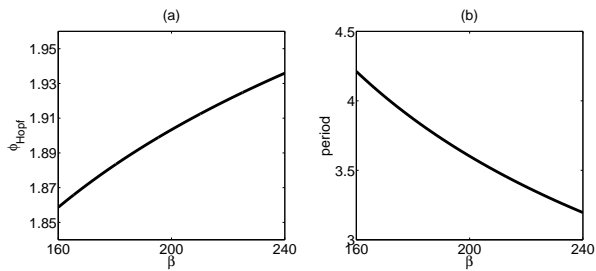}
  \end{center}
    \caption{Single patch behavior in the absence of cross immunity:
      (a) ADE value $\phi$ where the Hopf bifurcation occurs versus
      contact rate $\beta$. (b) Period of the periodic orbit versus
      $\beta$ for $\phi=1.903$ .  $\epsilon=0$, $\nu=0$, $\mu=0.02$,
      and other parameters as in Table \ref{tab1}.}\label{fig:1patch}
\end{figure}

Figure \ref{fig:1patch}a shows the bifurcation behavior of the single
patch model with no cross immunity as the contact rate $\beta$ is
varied.  The bifurcation structure was computed using a
continuation routine \citep{auto97}.  The ADE value at which the
bifurcation occurs increases slightly as $\beta$ increases, and the
dependence is approximately linear.  The period of the periodic orbit,
shown in Figure \ref{fig:1patch}b for a fixed ADE value, also varies
with $\beta$.  (Note that multistrain models with ADE can display
subcritical Hopf bifurcations \citep{BillingsJTB07}, so the periodic
orbit shown in Figure \ref{fig:1patch}b exists throughout the range of
$\beta$ values shown.)  Similarly, varying the contact rate $\beta$ in
the absence of ADE affects the location of the Hopf bifurcation in
cross immunity $\epsilon$ and its characteristic frequency of
oscillation.  Likewise, varying the other social parameter, the birth
rate $\mu$, affects the location and frequency of the Hopf
bifurcations in $\phi$ and $\epsilon$ (data not shown).

\section{Coupled patches}
\label{sec:2patch}

We next examine the effect of coupling between distinct regions on the
dynamics previously observed in the single patch model.  In
particular, we consider how migration between nonidentical patches
affects the stability of the steady state.  We investigate the
dynamics of the coupled systems by numerically integrating
Eqs.~\ref{odefirst}-\ref{odelast} and by tracking the bifurcations  using a
continuation routine \citep{auto97}.  As mentioned in the previous Section, the model has
a Hopf bifurcation at critical values of the bifurcation parameters $\phi$,
the ADE factor, and $\epsilon$, the cross
immunity strength.
We study the effect of coupling and asymmetry on the stability by observing their effect on the location of the Hopf bifurcations.
We consider cross immunity and ADE separately; that is, we analyze the system either with ADE and no cross
immunity or with cross immunity and no ADE.  Including both
cross immunity and ADE together leads to qualitatively similar behavior to
that reported here, at least locally when the asymmetry is not too large.

We proceed by fixing the patch-specific parameters in patch 2 and varying a social parameter ($\beta_1$ or $\mu_1$) in patch 1 to observe the effect of increasing asymmetry on the dynamics.   As previously mentioned,  changing the social
  parameters modifies the natural frequency of the system.  For each value of asymmetry, we
look for the critical points at which the coupled system loses
stability.  We also consider several  migration
rates $\nu$.

\begin{figure}[!ht]
  \begin{center}
    \includegraphics[width=9cm, keepaspectratio]{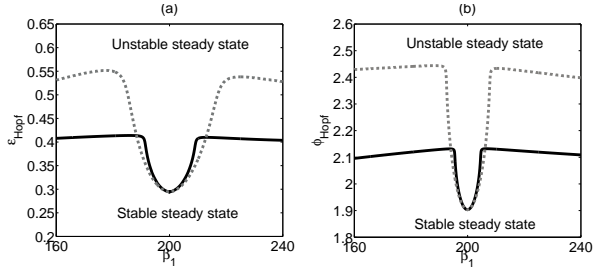}
    \caption{ Critical values of the parameters (a) $\epsilon$ and (b) $\phi$ at which the Hopf
      bifurcation occurs for coupled patches, as a function of the contact rate
      in patch $1$, for two values of the migration rate $\nu = 0.02$ (solid black)
      and $\nu = 0.05$ (dashed gray). The contact rate in patch $2$ is
      fixed at $\beta_2 = 200$.  In (a), $\phi=1$ (no ADE).  In (b), $\epsilon=0$ (no cross immunity).  $\mu_1=\mu_2=0.02$ and other parameters are as in Table \ref{tab1}. }\label{Asym_beta}
  \end{center}
\end{figure}

The effect of asymmetry in the contact rate is shown in Figs.~\ref{Asym_beta}(a)
and~\ref{Asym_beta}(b) for two migration rates, namely $\nu = 0.02$ (black) and $\nu
= 0.05$ (gray). The value of the critical parameter $\epsilon$
(Fig.~\ref{Asym_beta}(a)) or $\phi$ (Fig.~\ref{Asym_beta}(b)) at which
the Hopf bifurcation occurs is plotted against the varying contact rate. The
symmetric case is at the bottom of the curve. We see that two identical patches have the same bifurcation point as a single, well
mixed population even when migration is present (cf.~Fig.~\ref{fig:1patch}a). However, a striking difference in the dynamics appears
when we make the two patches weakly asymmetric. The values of $\phi$
and $\epsilon$ at which the Hopf bifurcation occurs are dramatically
different from the symmetric case. The steady state stability persists well into the parameter
regime where a single patch would display chaotic oscillations
\citep{BillingsJTB07,BiaShaSch09}.  The location of the single patch
Hopf point does not depend strongly on $\beta$, and decreasing the
contact rate is actually destabilizing (Figure \ref{fig:1patch}a), so
the stabilization observed here is clearly a result of the coupling
between asymmetric systems.
Slightly increasing the migration rate (from $\nu = 0.02$ to $\nu =
0.05$ in Figure \ref{Asym_beta}) further increases the stability of
the system.

Similar results are obtained if the
asymmetry occurs in the other social parameter, the birth rate.
Again, for asymmetric patches either stronger cross immunity
or stronger ADE is needed to destabilize the steady state through a
Hopf bifurcation (figure in Supplementary Material).

\section{Analytical motivation}
\label{sec:analytical}

We now motivate the results of the preceding Section using  a simple
lower dimensional model and show that the qualitative features depend
only on the bifurcation structure and characteristic frequencies
rather than on details of the epidemic model.  In this Section we
study the general behavior of two coupled systems, each displaying a
Hopf bifurcation, but with different characteristic frequencies.

The generic form for a Hopf bifurcation is the
following, in polar coordinates~\citep{Strogatz}:
\begin{equation}
  \label{Hopf}
      \begin{array}{rl}
        \dot{r} & = \lambda r - r^3\\
        \dot{\theta} & =  \omega 
        \end{array}
\end{equation}
A Hopf bifurcation occurs at $\lambda = 0$, where periodic oscillations with
frequency $\omega$ are observed. We now couple two systems of the sort in
Eq.~\ref{Hopf}, each with a potentially different frequency
$\omega_q$, via linear migration terms with migration rate $\nu$.  To
lowest order (not displaying cubic terms), the coupled system in
cartesian coordinates is
\begin{equation}
  \begin{array}{rl}
    \dot{x_1} & = \lambda x_1 - \omega_1 y_1 - \nu x_1 + \nu x_2 \\
    \dot{y_1} & = \lambda y_1 + \omega_1 x_1 - \nu y_1 + \nu y_2 \\
    \dot{x_2} & = \lambda x_2 - \omega_2 y_2 - \nu x_2 + \nu x_1 \\
    \dot{y_2} & = \lambda y_2 + \omega_2 x_2 - \nu y_2 + \nu y_1
  \end{array} \label{Hopfcoupled}
\end{equation}

Stability is determined by evaluating the Jacobian of Eq.~\ref{Hopfcoupled} at the steady state $(x_1,y_1,x_2,y_2)=0$.  At the Hopf bifurcation, the real part of the largest eigenvalue crosses zero, with nonzero imaginary part.  We obtain the Hopf point directly by setting the real part of the eigenvalue to zero and solving for the critical value of $\lambda$ at the bifurcation.  Details may be found in the Supplementary Material.  The Hopf bifurcation occurs at
\begin{equation}
\lambda_c =
\left\{
\begin{array}{ll}
\nu - \frac{1}{2} \sqrt{4 \nu^2 - (\omega_1-\omega_2)^2} & \text{ for } \left| \omega_1-\omega_2 \right| < 2 \nu\\
\nu & \text{ for } \left| \omega_1-\omega_2 \right| \geq 2 \nu
\end{array}
\right. \label{Hopfpoint}
\end{equation}

\begin{figure}[!ht]
  \begin{center}
    \includegraphics[width=8cm,height = 4cm]{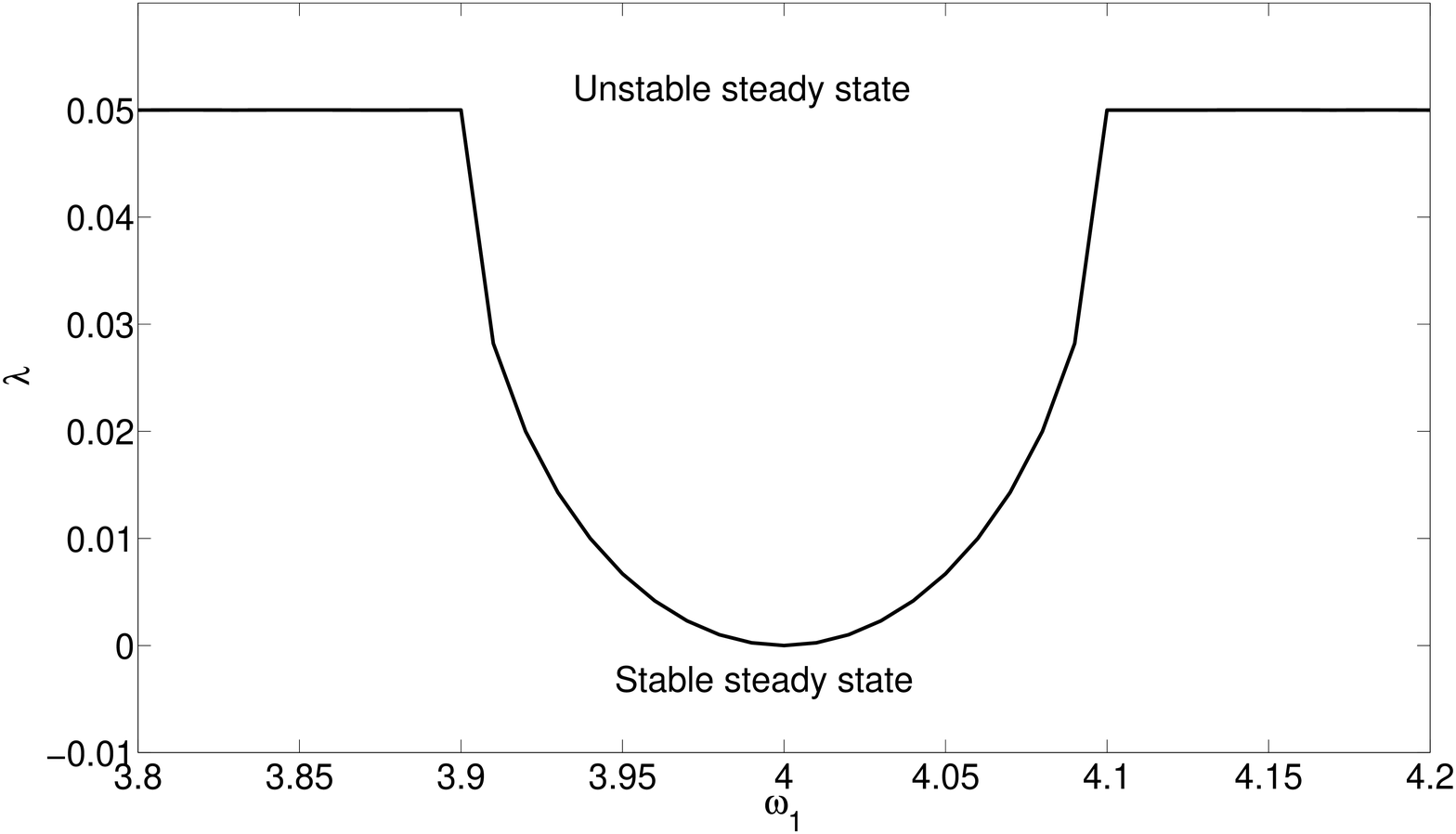}
  \end{center}
    \caption{Critical values of the parameter $\lambda$ at which the Hopf
      bifurcation occurs for the coupled generic Hopf bifurcations, as a
      function of the frequency $\omega_1$. Here the parameter $\omega_2$ has
      been kept fixed at $\omega_2=4$. $\nu = 0.05$.}\label{Asym_theo}
\end{figure}

In Fig.~\ref{Asym_theo} we show the location of the Hopf bifurcation given by Eq.~\ref{Hopfpoint}
as a function of the asymmetry between the two systems for $\nu =
0.05$. Comparison with Fig.~\ref{Asym_beta}
(and Fig.~1 of the Supplementary Material) shows the qualitative agreement of the theoretical
results for the lower dimensional system with the full multistrain
system. Increasing the migration rate $\nu$
increases the value of the bifurcation parameter to which the Hopf
point saturates (not shown), as in Fig.~\ref{Asym_beta}.
It is also worth noticing that, in the case of identical frequencies
$\omega_1 = \omega_2$, the Hopf bifurcation occurs at $\lambda_c
= 0$, the location in the absence of coupling.  This is consistent with the numerical results for the
multistrain system in the case of symmetric patches.

\section{Conclusions and discussion}
\label{sec:conclusions}
We have studied the endemic steady state stability properties for a
multistrain epidemic model on two migration-coupled patches.
Interactions between strains in the model were governed by temporary
partial cross immunity and antibody-dependent enhancement.  In the
absence of coupling, the system displayed Hopf bifurcations in two
epidemic parameters.  Coupling between patches with non-identical
parameters, which gave them non-identical characteristic frequencies
of oscillation, was shown to shift the Hopf bifurcations,
stabilizing the steady state.  This behavior was observed for the Hopf
bifurcation obtained by sweeping the cross immunity in the absence of
ADE and the bifurcation obtained by sweeping the ADE in the absence of
cross immunity.  It occurred for asymmetry in either of our two social
parameters, the birth rate and the contact rate.

To motivate this result, we diffusively coupled two low dimensional
Hopf bifurcations with different characteristic frequencies and
analyzed the stability of the steady state.  We again saw that
coupling between asymmetric systems led to stabilization.  This
indicates that the stabilization in the epidemic model is a result of
the underlying dynamics, rather than the details of the model.  The
stabilization may occur because when the frequencies of oscillation in
two coupled systems are different, the oscillations tend to cancel
each other because of phase differences.   This topic will be studied
in more detail in a future work.

Bifurcations from steady state to oscillatory behavior can be
associated with an increased number of infection cases, particularly
if chaotic oscillations occur, as in previous dengue models
\citep{BillingsJTB07,BiaShaSch09}.  Our results suggest that if
control strategies in one region are able to generate enough
asymmetry, this could lead to a stabilization of
the outbreaks, which would have a positive effect on adjacent regions.
Asymmetry could be generated in the effective contact rate by mosquito
control, which could include reducing mosquito breeding sites
\citep{Slosek1986}, or through new genetic controls which are under development
\citep{Barbazan2008}.  Asymmetry in the birth rate could be generated by
lowering the effective birth rate through vaccination of new susceptibles once
a vaccine is available.
However, because the bifurcation point saturates rather than
increasing indefinitely as asymmetry increases, such a strategy would
be successful only if the epidemic parameters in the real system are
moderately close to the bifurcation.  (Extreme asymmetry may even be
destabilizing, so this strategy works best locally when the asymmetry
is not too strong.) In addition, the role of
seasonality in exciting oscillations should not be ignored.  Seasonal
variations in the contact rate have been included in previous dengue
models \citep{LeahPRE}.  The interplay of seasonality and coupling is
a topic for future study.

The results of the present work may also be useful in
lower bound estimation of the parameter values for ADE and cross
immunity.  ADE especially is difficult to measure in vivo and must be
estimated by other means.  Since recent publications have
suggested that epidemiological data from Thailand show chaotic
outbreaks~\citep{LeahPRE}, and given the certainty of human migration
and asymmetry between adjacent regions of the country, it is
possible that the actual parameter values for ADE and cross immunity are higher
than the ones estimated by studying a single well mixed population.

Finally, the work discussed here shows a potential effect of human movement
between heterogeneous regions.  As spatial effects are further studied
in epidemic models, it remains to be seen how this phenomenon will
extend to more complicated spatial geometries, including more patches
and perhaps non-symmetric coupling terms.  Because the
migration-induced stabilization depends only on the existence of a
Hopf bifurcation in the model, it is expected that the stabilization
will be observed in other population models that also contain Hopf
bifurcations (e.g.,~\citet{Fussmann2000, Greenhalgh2004}).

\section{Acknowledgements}

This work was supported in part by the Jeffress Memorial Trust.  The
authors wish to thank Ira Schwartz for helpful discussions.


\appendix


\section{Asymmetric birth rates}

\begin{figure}[tbp]
  \begin{center}
    \includegraphics[width=9cm, keepaspectratio]{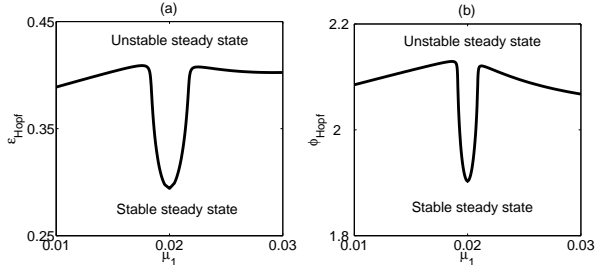}
    \caption{ Critical values of the parameters (a) $\epsilon$ and (b) $\phi$ at which the Hopf
      bifurcation occurs for coupled patches, as a function of the birth/death rate
      in patch $1$, for  $\nu = 0.02$. The birth/death rate in patch $2$ is
      fixed at $\mu_2 = 0.02$.  In (a), $\phi=1$ (no ADE).  In (b), $\epsilon=0$ (no cross immunity).  $\beta_1=\beta_2=200$ and other parameters are as in Table I of the paper.}\label{Asym_mu}
  \end{center}
\end{figure}

The effect of asymmetric birth rates is shown in Fig.~\ref{Asym_mu}
 of this supplementary material.
The birth rate in patch 2 was held fixed while that in patch 1 was
varied.  For asymmetric patches either stronger cross immunity
or stronger ADE is needed to destabilize the steady state through a
Hopf bifurcation, compared to the symmetric patch case.

\section{Details of Hopf bifurcation analysis}

We evaluate the Jacobian for the ODE system given in Eq.~7 of the
paper at the steady state $\{x_1,y_1,x_2,y_2\}=0$.  The roots of the
characteristic polynomial $f(t)$ of the 
Jacobian are the eigenvalues $\{ t \}$.  The four eigenvalues are 
  \begin{eqnarray}
    t_{1/2} & = & \lambda - \nu + \frac{1}{2}\sqrt{4 \nu^2 - 2(\omega_1^2 +
      \omega_2^2) \pm 2 (\omega_1 + \omega_2)\sqrt{(\omega_1 - \omega_2)^2 - 4\nu^2}}\label{longEig1}\\
    t_{3/4} & = & \lambda - \nu - \frac{1}{2}\sqrt{4 \nu^2 - 2(\omega_1^2 +
      \omega_2^2) \pm 2 (\omega_1 + \omega_2)\sqrt{(\omega_1 - \omega_2)^2 - 4\nu^2},}\label{longEig2}
  \end{eqnarray}
but it is not easy to see from these expressions where the Hopf bifurcation occurs.  Instead, we solve directly for the Hopf point.

At a Hopf bifurcation, a pair
of eigenvalues has zero real part.  Thus to obtain the Hopf point, we
can set $t=i\mu$, where $\mu$ is real.
If $i \mu$ is an eigenvalue, we require $f(i\mu)=0$, so
\begin{equation}
\Re [f(i\mu)]=0 \label{realpart}
\end{equation}
and
\begin{equation}
\Im [f(i\mu)]=0 \label{imagpart}
\end{equation}

If $\lambda \neq \nu$, Eq.~\ref{imagpart} has nonzero roots $\mu = \pm
\frac{1}{2} \sqrt{4 \lambda^2 - 8 \lambda \nu +2 \omega_1^2 +2
  \omega_2^2}$.  When this $\mu$ is substituted into
Eq.~\ref{realpart}, we obtain four potential roots for $\lambda$ at
the Hopf point.  Two are complex, which is unphysical and can be
ignored.  The one corresponding to the Hopf bifurcation is $\lambda_c
= \nu - \frac{1}{2} \sqrt{4 \nu^2 - (\omega_1-\omega_2)^2}$.  (The
fourth root is larger and corresponds to loss of stability of the
second pair of eigenvalues.)  This $\lambda_c$ is real and thus gives
the Hopf point location when $\left| \omega_1-\omega_2 \right| \leq 2
\nu$.  When $\omega_1-\omega_2 \neq 0$ and $\nu>0$, $\lambda_c$ is positive,
indicating stabilization due to the asymmetry and coupling.

When $\lambda = \nu$, Eq.~\ref{imagpart} is always satisfied.  In that
case, we must turn to Eq.~\ref{realpart} to determine $\mu$.
Eq. \ref{realpart} has four roots, $\mu = \pm \frac{1}{2}
(\omega_1+\omega_2) \pm \frac{1}{2} \sqrt{(\omega_1-\omega_2)^2-4
  \nu^2}$.  $\mu$ must be real by our assumption, and this occurs when
$\left| \omega_1-\omega_2 \right| \geq 2 \nu$, which is precisely the
case not covered by the above result.  Thus when $\left| 
\omega_1-\omega_2 \right| \geq 2 \nu$, the Hopf point is at $\lambda_c = \nu$.

\end{document}